\documentclass{aa-mod} 
\usepackage{graphicx}
\usepackage{natbib}
\usepackage{txfonts}
\usepackage{longtable}

\newcommand{\hess}{H\textsc{.E.S.S.}}
\newcommand{\gr}{\ensuremath{\gamma}-ray}
\newcommand{\grs}{\ensuremath{\gamma}-rays}
\newcommand{\dg}{\ensuremath{^{\circ}}}
\newcommand{\on}{\emph{on}}
\newcommand{\off}{\emph{off}}

\newcommand{\iflux}{\ensuremath{\textrm{cm}^{-2} \textrm{s}^{-1}}}
\newcommand{\stat}{\ensuremath{_{\textrm{stat}}}}
\newcommand{\sys}{\ensuremath{_{\textrm{sys}}}}
\newcommand{\chisq}{\ensuremath{\chi^{2}}}
\newcommand{\rchisq}{\ensuremath{\chi^{2}}/dof}

\begin{document}
\renewcommand{\topfraction}{0.85}
\renewcommand{\bottomfraction}{0.7}
\renewcommand{\textfraction}{0.15}
\renewcommand{\floatpagefraction}{0.66}

\title{First detection of a VHE gamma-ray spectral maximum from a Cosmic source:
  \hess\ discovery of the Vela\,X nebula.}

\titlerunning{First detection of a VHE gamma-ray spectral maximum from a Cosmic source}

\author{F. Aharonian\inst{1}
 \and A.G.~Akhperjanian \inst{2}
 \and A.R.~Bazer-Bachi \inst{3}
 \and M.~Beilicke \inst{4}
 \and W.~Benbow \inst{1}
 \and D.~Berge \inst{1}
 \and K.~Bernl\"ohr \inst{1,5}
 \and C.~Boisson \inst{6}
 \and O.~Bolz \inst{1}
 \and V.~Borrel \inst{3}
% ...
}

\institute{
Max-Planck-Institut f\"ur Kernphysik, P.O. Box 103980, D 69029
Heidelberg, Germany
\and
 Yerevan Physics Institute, 2 Alikhanian Brothers St., 375036 Yerevan,
Armenia
\and
Centre d'Etude Spatiale des Rayonnements, CNRS/UPS, 9 av. du Colonel Roche, BP
4346, F-31029 Toulouse Cedex 4, France
\and
Universit\"at Hamburg, Institut f\"ur Experimentalphysik, Luruper Chaussee
149, D 22761 Hamburg, Germany
\and
Institut f\"ur Physik, Humboldt-Universit\"at zu Berlin, Newtonstr. 15,
D 12489 Berlin, Germany
\and
LUTH, UMR 8102 du CNRS, Observatoire de Paris, Section de Meudon, F-92195 Meudon Cedex,
France
% ...
}

\abstract{The Vela supernova remnant (SNR) is a complex region
  containing a number of sources of non-thermal radiation. The inner
  section of this SNR, within 2 degrees of the pulsar \object{PSR B0833$-$45},
  has been observed by the \hess\ \gr\ atmospheric Cherenkov detector
  in 2004 and 2005. A strong signal is seen from an extended region to
  the south of the pulsar, within an integration region of radius
  $0.8\dg$ around the position ($\alpha = 08^{h} 35^{m} 00^{s}$,
  $\delta\ = -45\dg\ 36\arcmin$ J2000.0).  The excess coincides with a
  region of hard X-ray emission seen by the ROSAT and ASCA satellites.
  The observed energy spectrum of the source between 550 GeV and 65
  TeV is well fit by a power law function with photon index $\Gamma =
  1.45 \pm 0.09\stat\ \pm 0.2\sys$ and an exponential cutoff at an
  energy of $13.8 \pm 2.3\stat\ \pm 4.1\sys$ TeV. The integral flux
  above 1 TeV is $(1.28 \pm 0.17\stat\ \pm 0.38\sys) \times 10^{-11}\ 
  \iflux$.  This result is the first clear measurement of a peak in
  the spectral energy distribution from a VHE \gr\ source, likely
  related to inverse Compton emission. A fit of an Inverse Compton
  model to the \hess\ spectral energy distribution gives a total
  energy in non-thermal electrons of $\sim$$2 \times 10^{45}$ erg
  between 5 TeV and 100 TeV, assuming a distance of 290 parsec to the
  pulsar. The best fit electron power law index is $2.0$, with a
  spectral break at 67 TeV.
  
}

\maketitle

\section{Introduction}

The region surrounding the Vela supernova remnant (SNR) is very well
studied across the electromagnetic spectrum and contains a number of
complex objects, including the SNR \object{RX J0852.0$-$4622}, which has been
previously detected in the very high energy \gr\ (VHE) range
\citep{katagiri05,aharonian05b}.  The Vela SNR itself, at a distance
estimated to be $\sim$290~pc \citep{dodson03,caraveo01}, extends over
a diameter of $\sim$$8\dg$. It is the nearest SNR to contain a young
active pulsar, \object{PSR B0833$-$45}, with a period of 89 ms and a period
derivative \.P of $1.25 \times 10^{-13}\ \textrm{s}\ \textrm{s}^{-1}$.
This implies a spin-down luminosity of $7 \times 10^{36}\ 
\textrm{erg}\ \textrm{s}^{-1}$ and age of 11,000 years.  Observations
by Chandra \citep{helfand01} clearly show the torus-like morphology of
the compact X-ray nebula surrounding the pulsar and have allowed its
rotation axis to be inferred. The plane of the torus (i.e. pulsar
equator) appears to intersect the plane of the sky at a position angle
of $40.6\dg$ relative to North \citep{ng04}, and its X-ray spectral
index of $\sim$1.5 \citep{mangano05} is typical of pre-cooled pulsar
wind torii.

The SNR also contains a number of regions of non-thermal emission,
including those labelled by \citet{rishbeth58} from radio observations
as Vela X, Vela Y and Vela Z (which is part of the shell of \object{RX
J0852.0$-$4622}).  A diffuse emission feature has been detected by
ROSAT \citep{markwardt95} in hard X-rays ($0.9$--$2.0$~keV),
coinciding with the centre of the Vela X region.  It was first
suggested that this feature, which is aligned closely with a filament
detected at radio wavelengths, corresponds to the outflow jet from the
pole of the pulsar \citep{frail97}. However, the Chandra observations
showed that this feature lies along the extension of the pulsar
equator, although bending to the southwest.

The study of Vela X is important from two perspectives: It was the
first middle-aged pulsar wind nebula (PWN) to be detected, thereby
introducing the concept of PWN evolution \citep{weiler80}. Also, it
served as the first prototype for offset PWN as a result of SNR
expansion into an inhomogeneous medium \citep{blondin01}, of which
\object{G18.0-0.7} served as a second example \citep{gaensler03}. This second
PWN was also recently identified with VHE source \object{HESS J1825--137}
\citep{aharonian05c}.

Here we discuss observations of the Vela pulsar and the centre of the
Vela SNR with \hess, including the Vela~X feature. A region of
extended \gr\ emission is detected, to the south of the pulsar
position (previously reported by \citet{khelifi05}.  This region was
previously observed by the CANGAROO collaboration \citep{yoshikoshi97}
and emission of VHE \grs\ was claimed from a source $0.13\dg$ to the
southeast of the pulsar. However doubt was later cast on the
significance of this detection \citep{dazeley01}. A recent paper from
the CANGAROO collaboration \citep{enomoto05} retracts the claimed
detection, and suggests some evidence for an excess coincident with
the \hess\ signal.

\section {Observations with \hess}
   
\hess\ is a system of four atmospheric Cherenkov telescopes designed
to look for astrophysical \gr\ emission above $\sim$100 GeV; a review
of the detector is given by \citet{hinton04}. The Vela region was
observed between January and March 2004 with the complete \hess\ 
array. In total 10.3 hours of data were obtained, after standard data
quality selection, at zenith angles between 20\dg\ and 40\dg. These
observations were taken using a method (\emph{wobble} mode) whereby
the source is offset by a small angular distance from the centre of
the field of view, alternating between 28 minute runs in the positive
and negative declination (or right ascension) directions.  The
observations were made at offsets of 0.5\dg\ in declination from the
position of the Vela pulsar, ($\alpha = 08^{h} 35^{m} 20^{s}$,
$\delta\ = -45\dg\ 10'\ 35''$ J2000.0), which is referred to as
position I for the purposes of this paper.

Following a detection of extended \gr\ emission to the south of the
pulsar in the 2004 dataset, further observations were made in 2005
surrounding the position $\alpha = 08^{h} 35^{m} 00^{s}$, $\delta\ =
-45\dg\ 36'$ (J2000.0), here referred to as position II, which was
measured as the centre of gravity of the excess. These were also taken
in wobble mode, and the mean offset for the 2005 data is 0.5\dg. After
data quality selection 7.2 hours of observations were available for
analysis, giving total observation live time over the two years of
16.4 hours, after dead time correction (approximately 10\%). The mean
zenith angle for the complete set of observations is $30.2\dg$.

\section{Analysis}

The observations of the Vela region were calibrated using standard
\hess\ calibration procedures, as discussed by \cite{aharonian04b} and
analysed using the \emph{scaled width} method, as discussed by
\cite{aharonian05a}. The \textit{hard} cuts, as defined in the latter
paper were applied to this source. These cuts improve the angular
resolution and reduce systematic effects due to uncertainties in the
background estimation, relative to the standard cuts, albeit at the
expense of a higher than usual energy threshold (450 GeV for the analysis
presented here).

The background level in this analysis was estimated using the
\emph{on-off} method, based on a sample of runs taken at similar
zenith angles (external to the Vela SNR) which contain no excess \gr\ 
signal. This method allows the background level to be estimated in
cases where the integration region for the source is comparable to the
size of the field of view. The off runs are carefully chosen to match
well the on source observations in zenith angle and system
configuration.

\begin{figure}
  \centering
  \includegraphics[width=0.5\textwidth]{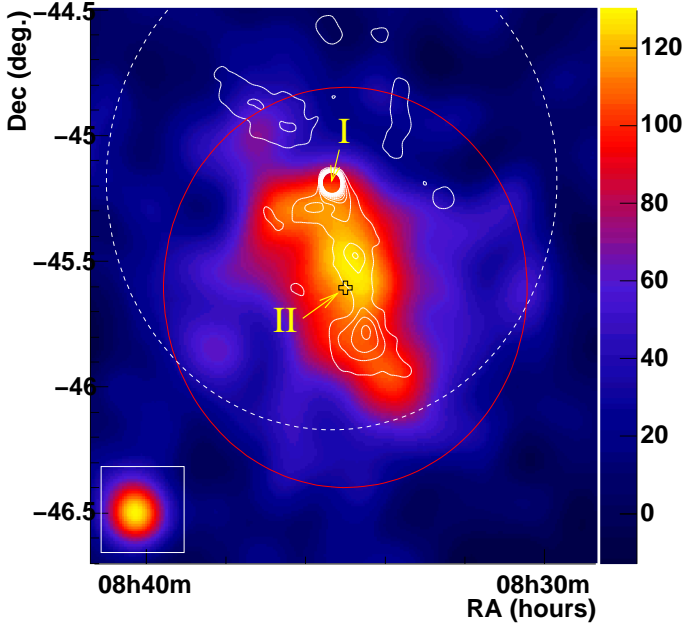}
  \caption{Gaussian smoothed sky map of region surrounding Vela pulsar,
    showing significant emission to the south of the pulsar position,
    coincident with an X-ray feature seen by ROSAT (white contours).
    The smoothing width used is 0.09\dg. The contours corresponding to
    the strong emission close to the pulsar (Position I) are
    truncated. The image inset in the bottom left corner indicates the
    size of a point source as seen by \hess, for an equivalent
    analysis. The solid circle represents the \hess\ integration region
    for the spectral measurement, while the dashed circle represents
    the field of view for the ROSAT observations. Position II is
    marked by a black cross.}
  \label{fig:skymap_narrow}
\end{figure}

Figure \ref{fig:skymap_narrow} shows a sky map of the excess in the
region south of the Vela pulsar. The map has been smoothed with a
Gaussian function with a sigma of $0.09\dg$ and is not corrected for
variations in the \gr\ radial acceptance of the detector, which are less
than 10\% across the integration region. Position II, as defined
above, is marked by a black cross. The extent and position of the
excess has been determined by fitting a two dimensional, elongated
Gaussian, convolved with the point spread function of the instrument
for these measurements, which is 0.08\dg\ (68\% containment radius),
to an uncorrelated excess map of the source. The best fit intrinsic
width along the major axis is $0.48\dg \pm 0.03\dg$, while the best
fit intrinsic width along the minor axis is $0.36\dg \pm 0.03\dg$.
The major axis of the fitted distribution is at a position angle of
$41\dg \pm 7\dg$. The best-fit centre of gravity of the emission
region ($\alpha = 08^{h} 35^{m} 1^{s}$, $\delta\ = -45\dg\ 34'\ 40''$ J2000.0)
is consistent with position II, within the statistical error of 2
arcminutes. The VHE source is thus identified as \object{HESS J0835--455}.

\begin{figure}
  \centering
  \includegraphics[width=0.45\textwidth]{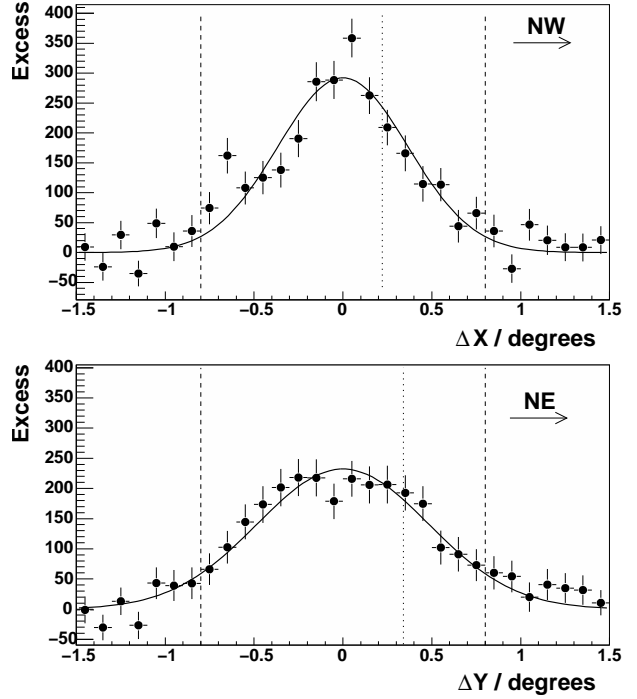}
\caption{Profile of the excess from the extended feature coincident with Vela X
  along minor axis (top) and along major axis (bottom); events within
  $0.8\dg$ of the axis are integrated into the profile in each case.
  The two ~dimensional fit of an elongated Gaussian convolved with the
  \hess\ PSF is also shown as a profile for comparison. The position
  of the Vela pulsar is marked by a vertical dotted line in each plot
  for reference, while the integration region in each case for the
  spectral measurement is marked by dashed lines (within $0.8\dg$ of
  position II).}
\label{fig:slices}
\end{figure}

Profiles of the excess parallel to, and perpendicular to, the major
axis of the fit are shown in Figure \ref{fig:slices}. Profiles of the
elongated Gaussian fit are shown for comparison. It can be seen that,
given the statistical errors in each bin, the elongated Gaussian fits
the excess well, with a \chisq\ of 277.8 for 250 degrees of freedom. A
circular Gaussian fit to the same excess gives a width of $0.43\dg \pm
0.02\dg$, with a \rchisq\ of 290.0/252.

The radius of the integration region used for estimating the flux and
spectrum of the VHE emission was $0.8\dg$, surrounding position II, as
defined above. This size was chosen so as to encompass the entire
excess, as seen in Figure \ref{fig:skymap_narrow}. A total of 9610
\on-source counts and 7879 \off-source counts were detected above 450
GeV, with a normalisation factor of 0.94 between the \on\ and \off\ 
observations, giving an excess of 2152 events within the integration
region. The statistical excess significance is $16.7\sigma$,
calculated using the method described by \citet{li83}.

Figure \ref{fig:spectrum} shows the energy spectrum of the signal seen
from the integration region. The spectrum is well fit (with a \rchisq\ 
of 13.1/10) by a power law function with photon index $\Gamma = 1.45
\pm 0.09\stat\ \pm 0.2\sys$ and an exponential cutoff at an energy of
$13.8 \pm 2.3\stat\ \pm 4.1\sys$ TeV. The data may alternatively be
described using a broken power law function, with a spectral break at
$13.4 \pm 3.1\stat\ \pm 4.1\sys$ TeV and photon index of $\Gamma_{1} =
1.7 \pm 0.1\stat\ \pm 0.2\sys$ at lower energies hardening to
$\Gamma_{2} = 3.4 \pm 0.7\stat\ \pm 0.2\sys$ above the break energy.
The \rchisq\ of the broken power law fit, which includes a term for
the width of the transition region, is 9.0/8. A straight power law fit
(\rchisq\ of 74.1/11) is rejected.  The integral flux above 1 TeV
(from the exponential cutoff fit) is $(1.28 \pm 0.17\stat\ \pm
0.38\sys) \times 10^{-11}\ \iflux$. The systematic error on the flux
estimation is estimated at 30\%, mainly due to uncertainties in the
transparency of the atmosphere to Cherenkov light. In fitting the
energy spectrum for this source, the energy of each event is corrected
for the detector optical efficiency, relative to that used in Monte
Carlo simulations to estimate the effective area of the instrument.
The optical efficiency is estimated from single muon events detected
during each observation run \citep{leroy03,bolz04}. The mean energy
correction is $\sim$$20\%$.

In order to test for a point source component in VHE \grs\ from
position I (the pulsar position, as marked in Figure
\ref{fig:slices}), the residual excess within 0.1\dg\ from the pulsar,
after subtraction of the Gaussian fit to the extended excess, has been
measured. No significant excess after subtraction is seen, and a
$99.9\%$ upper limit on the integral flux above 1 TeV, assuming a
point source at the position of the pulsar, is $7.6 \times 10^{-13}\ 
\iflux$.

\begin{figure}
  \centering
  \includegraphics[width=0.5\textwidth, height=10cm]{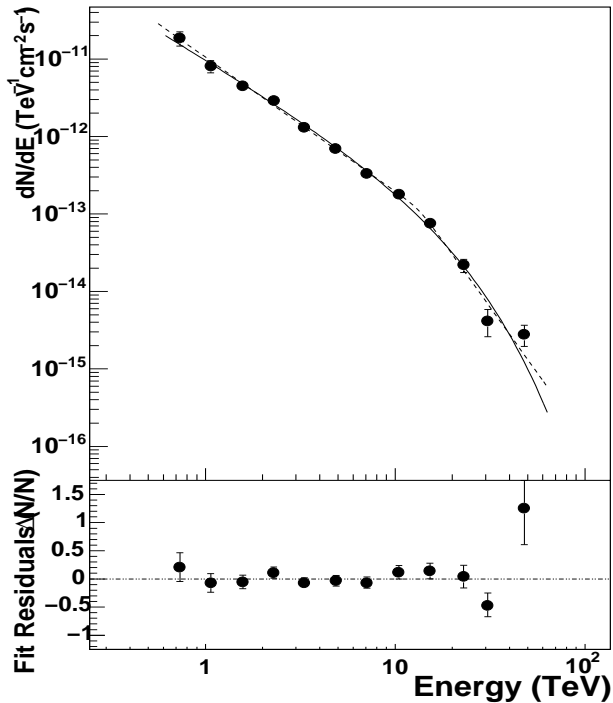}
  \caption{Energy spectrum of \gr\ emission from the Vela X region. 
    The solid line denotes the best fit of a power law with an
    exponential cutoff. The dashed line represents the best fit broken
    power law spectrum. The bottom panel shows the residuals to
    the exponential cutoff fit.}
  \label{fig:spectrum}
\end{figure}

\section{Discussion}

The new VHE source reported here, \object{HESS J0835--455}, is situated to the
south of the pulsar and the compact X-ray nebula (as seen by Chandra).
The integral flux is estimated to be $\sim$50$\%$ of that of the Crab
nebula above 1 TeV.  As the distance to the pulsar is well measured,
one can estimate the size of the emission region seen by \hess\ to be
5.1 parsec (full length for 68\% containment) along the major axis by
3.8 parsec (full width).  The luminosity of the emission region in the
energy range from 550 GeV to 65 TeV can be estimated to be $L =
9.9 \times 10^{32}\ \textrm{erg}\ \textrm{s}^{-1}$, using the power
law fit to the spectrum with the exponential cutoff.

\object{HESS J0835--455} appears to be spatially coincident with the X-ray
($0.4$--$2.4$ keV) emission as seen by ROSAT (shown in Figure
\ref{fig:skymap_narrow}). It has been suggested \citep{blondin01} that
the Vela X feature corresponds to the pulsar wind nebula, displaced to
the south by the unequal pressure of the reverse shock from the SNR.
This hypothesis is consistent with the \hess\ observations which
demonstrate conclusively for the first time that this feature emits
non-thermal radiation. A similar explanation has been proposed for the
emission seen by \hess\ (\object{HESS J1825--137}) close to the pulsar wind nebula G18.0--0.7
\citep{aharonian05c}.

The size of the feature described by \citet{markwardt95} has been
measured as 45 $\times$ 12 arcmin$^{2}$, whereas the VHE intrinsic
size is 58 $\times$ 43 arcmin$^{2}$. An extension of the X-ray feature
to the southwest was suggested by \citet{lu00}, however this may be
unrelated thermal emission, a suggestion which is supported by the
fact that no excess is seen in this region by \hess

\begin{figure}
  \centering
  \includegraphics[width=0.5\textwidth, height=7cm]{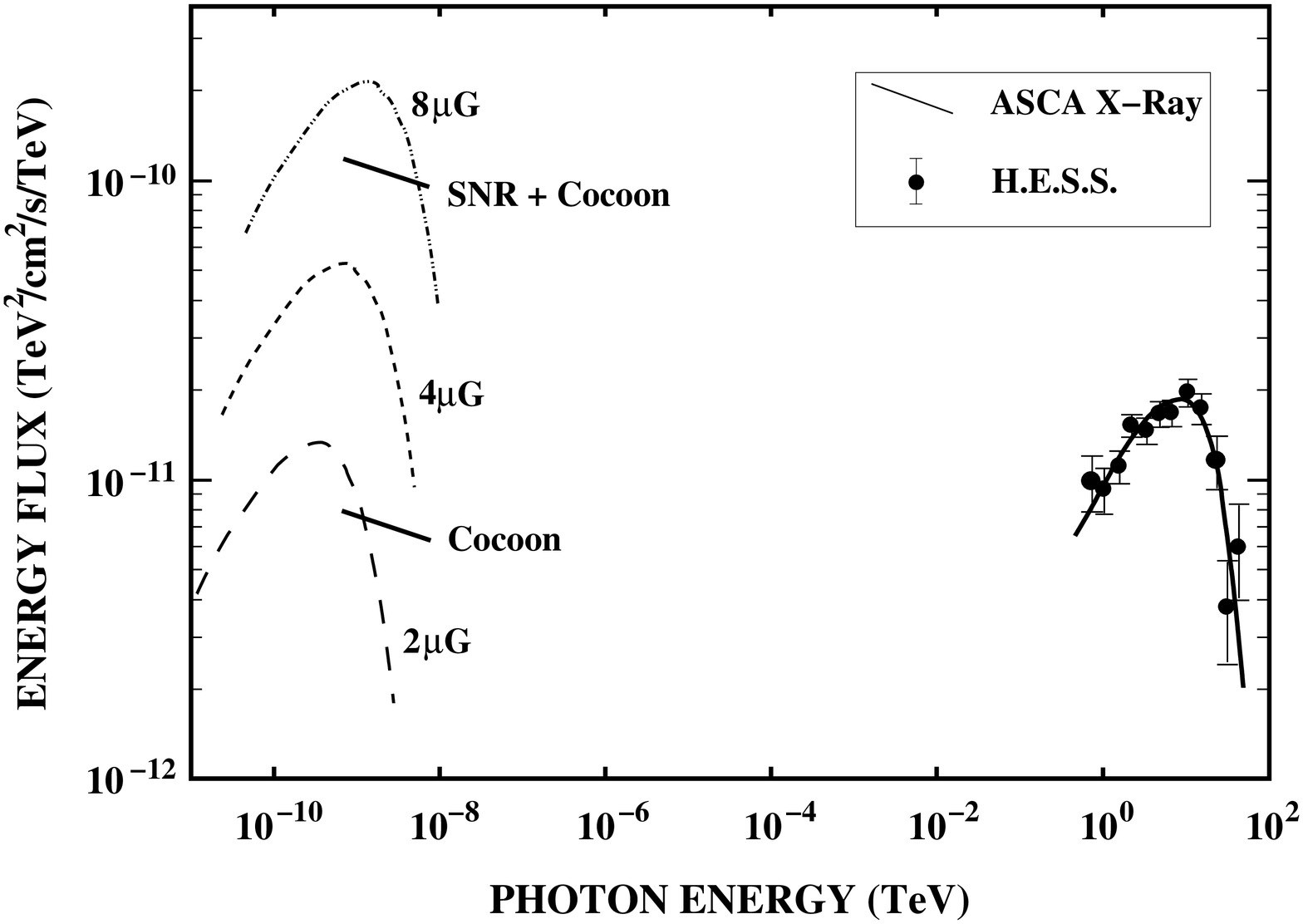}
  \caption{Spectral energy distribution for the \hess\ and ASCA spectral measurements
    \citep{markwardt97}. The two alternative X-ray spectra are described
    in the text.  The fitted inverse Compton emission (solid line)
    from the Vela X region is shown, given the electron energy
    distribution described in the text. The predicted synchrotron flux
    is shown for three possible magnetic field levels, between
    $2\ \mu$G and $8\ \mu$G.}
  \label{fig:sed}
\end{figure}

The larger VHE size probably indicates that synchrotron cooling in the
X-ray domain was (or still is) important.  This is confirmed by the
detection of a cooled X-ray photon index of $\sim$2.1
\citep{markwardt97} in the emission region, which is consistent with
the uncooled photon index of 1.5 closer to the source. Figure
\ref{fig:sed} shows a spectral energy distribution (SED) for Vela X,
including the \hess\ result reported here. Two possible X-ray spectra
are shown, the lower including only the 'Cocoon' feature defined in
the above paper, while the upper flux includes the emission from the
'SNR' component extrapolated over the \hess\ integration region.

A simple one-zone model of inverse Compton emission from interactions
with the cosmic microwave background (CMBR) is also shown.  The \hess\ 
observations show the expected peak in the IC emission; this result is
the first clear measurement of such a peak at VHE energies. The total
energy content in non-thermal electrons between 5 TeV and 100 TeV is
$2.2 \times 10^{45}$ erg, for a distance of 290 parsec.  The best fit
electron power law index is $2.0$, with a break at 67 TeV and a
post-break index of 9. The \rchisq\ of this fit is 10.2/9.  In Figure
\ref{fig:sed} three possible synchrotron spectral energy distributions
are shown, for magnetic field levels in the emission region from $2\ 
\mu$G to $8\ \mu$G. The synchrotron distributions are not constrained
by the X-ray data, a combined fit on the X-ray and \gr\ data would
more closely reproduce the measured X-ray spectral slope. A magnetic
field of this magnitude may be expected in the scenario of a nebula
displaced from the pulsar by an asymmetric shock.

This result demonstrates the usefulness of \hess\ observations in
clarifying the complex morphology of this source. VHE observations of
inverse Compton scattering of the CMBR allow direct inference of the
spatial and spectral distribution of non-thermal electrons in a PWN,
independent of contamination by thermal emission or variations in the
local magnetic field.

\begin{acknowledgements}
  The support of the Namibian authorities and of the University of
  Namibia in facilitating the construction and operation of H.E.S.S.
  is gratefully acknowledged, as is the support by the German Ministry
  for Education and Research (BMBF), the Max Planck Society, the
  French Ministry for Research, the CNRS-IN2P3 and the Astroparticle
  Interdisciplinary Programme of the CNRS, the U.K. Particle Physics
  and Astronomy Research Council (PPARC), the IPNP of the Charles
  University, the South African Department of Science and Technology
  and National Research Foundation, and by the University of Namibia.
  We appreciate the excellent work of the technical support staff in
  Berlin, Durham, Hamburg, Heidelberg, Palaiseau, Paris, Saclay, and
  in Namibia in the construction and operation of the equipment. We
  have made use of the ROSAT Data Archive of the Max-Planck-Institut
  f\"ur Extraterrestrische Physik (MPE) at Garching, Germany.
\end{acknowledgements}

\bibliographystyle{aa}

\end{document}